\documentstyle[12pt,twoside,fleqn,espcrc1]{article}

\newcommand{\AmS}{{\protect\the\textfont2A\kern-.1667em
\lower.5ex\hbox{M}\kern-.125emS}} 
\newcommand {\be} {\begin{eqnarray}} 
\newcommand {\ee} {\end{eqnarray}} 
\begin{document} 
 \topmargin -0.3in
\oddsidemargin -0.50cm 
\evensidemargin 0cm 
\textwidth 6.5in
\textheight 8.5in 
\parindent 1.2cm 
\renewcommand{\textfraction} {.01} 
\renewcommand{\topfraction} {.99} 
\input psfig.sty
 
\pagestyle{empty}
\Huge{\noindent{Istituto\\Nazionale\\Fisica\\Nucleare}}

\vspace{-3.9cm}

\Large{\rightline{Sezione di ROMA}}
\normalsize{}
\rightline{Piazzale Aldo  Moro, 2}
\rightline{I-00185 Roma, Italy}

\vspace{0.65cm}

\rightline{INFN-1262/99}
\rightline{September 1999}

\vspace{1.cm}

\begin{center}{\large\bf A Poincar\'e-covariant current operator for
interacting systems and deuteron electromagnetic form factors}
\end{center}
\vskip 1em
\begin{center} F.M. Lev$^a$,  E. Pace$^b$ and G. Salm\`e$^c$ \end{center}

\noindent{$^a$\it Laboratory of Nuclear Problems, Joint Institute
for Nuclear Research, Dubna, Moscow region 141980, Russia}

\noindent{$^b$\it Dipartimento di Fisica, Universit\`a di Roma
"Tor Vergata", and Istituto Nazionale di Fisica Nucleare, Sezione
Tor Vergata, Via della Ricerca Scientifica 1, I-00133, Rome,
Italy}

\noindent{$^c$\it Istituto Nazionale di Fisica Nucleare, Sezione
di  Roma I, P.le A. Moro 2, I-00185 Rome, Italy}

\vspace{2.cm}

\begin{abstract}
In front-form dynamics a current operator for systems of interacting particles, which fulfills Poincar\'e, parity and time reversal   covariance,
together with hermiticity, can be defined. The electromagnetic form factors
can be extracted without any ambiguity and in the elastic case the continuity
equation is automatically satisfied. Applications to the calculation
of deuteron form factors are presented, and the effects of different nucleon-nucleon interactions, as well as of
different nucleon form factors are investigated.
\end{abstract}

\vspace{5.5cm}
\hrule width5cm
\vspace{.2cm}
\noindent{\normalsize{Proceedings of PANIC 99,  Uppsala, June 1999. 
To appear in {\bf Nucl. Phys. A}}} 

\newpage


\pagestyle{plain}
\section{INTRODUCTION}
\label{S1}
The electromagnetic (e.m.) current
operator and the states of a system must have correct transformation
properties with respect to the {\em same} representation of the Poincar\'e
group. For systems of interacting particles some of the Poincar\'e
generators are interaction dependent and, therefore, it is not a simple task to
construct a current operator, able to fulfill the proper Poincar\'e,
parity and time reversal covariance, as well as current conservation,
hermiticity and charge normalization. Instead of including perturbatively the
relativistic properties through $p/E$ series, or considering field-theoretic
approaches, we adopt the front-form Hamiltonian dynamics \cite{Dir} with a fixed number of particles, which
includes relativity in a coherent, non-perturbative way and allows one to
retain a large amount of the successfull phenomenology developed within the
"non-relativistic" domain. In this dynamics seven, out of ten,
Poincar\'e generators are interaction free. In particular, the Lorentz boosts
are interaction free and the states can be written as products of total momentum
eigenstates times intrinsic eigenstates.
 In the two-body case, if the mass operator, ${\tilde M}$, for the intrinsic
functions is defined according to ${\tilde M}^2 = M_0^2+V$ (with $M_0$
the free mass operator and $V$ the interaction operator), then the mass
equation has the same form as the nonrelativistic Schroedinger equation in
momentum representation \cite{Coe}.

In Ref. \cite{LPS}(a) it was shown that all the requirements
of Poincar\'e covariance can be satisfied, if, in the Breit frame where
the momentum transfer, $\vec q$, is directed along the spin-quantization axis,
$z$, the current operator is covariant with respect to rotations around $z$.
Since in the front form the rotations around the $z$ axis are kinematical, the
extended Poincar\'e covariance (i.e., 
Poincar\'e covariance plus parity and time reversal covariance) is satisfied by a current
operator which in our Breit frame is given by the sum of free, one-body currents
(i.e., $J^{\mu}(0) = \sum_{i=1}^{N} j_{free,i}^{\mu}$, with N the number of
constituents in the system). The hermiticity can be easily 
implemented and, in the elastic case, the extended Poincar\'e covariance plus
hermiticity imply current conservation \cite{LPS}(a,e). 
 
  Our Poincar\'e covariant current operator has been already succesfully tested
in the case of deep inelastic scattering \cite{LPS}(b). In this paper we
calculate  deuteron e.m. properties; more details for the magnetic moment
and the quadrupole moment and other results for the deuteron elastic form
factors (f.f.) can be found in Refs. \cite{LPS}(e) and \cite{LPS}(c,d),
respectively.

\section{DEUTERON ELECTROMAGNETIC FORM FACTORS}
\label{S2}
In the elastic case, for a system of spin $S$ one has only $2S+1$ non-zero
independent matrix elements for the current defined in Ref. \cite{LPS}(a),
corresponding to the $2S+1$ elastic form factors.
Then the extraction of elastic e.m. form factors is no more plagued by the
ambiguities which are present when, as usual, the free current is considered in
the reference frame where $q^+ = q_o + q_z = 0$ (indeed, if the current is taken
free in the $q^+=0$ frame, one has four independent matrix
elements to calculate the three deuteron f.f. \cite{GrKo}).

 Our results for the deuteron magnetic and quadrupole moments,
corresponding to different $N-N$ interactions, are reported in Fig. 1, together
with the non relativistic ones, against the deuteron asymptotic normalization
ratio $\eta = A_D/A_S$. A remarkable linear behaviour appears for both
quantities and in our Poincar\'e covariant calculation the relativistic effects
bring both $\mu_d$ and $Q_d$ closer to the experimental values, except for the
charge-dependent Bonn interaction \cite{CDB}.
\begin{figure}[t]
\psfig{figure=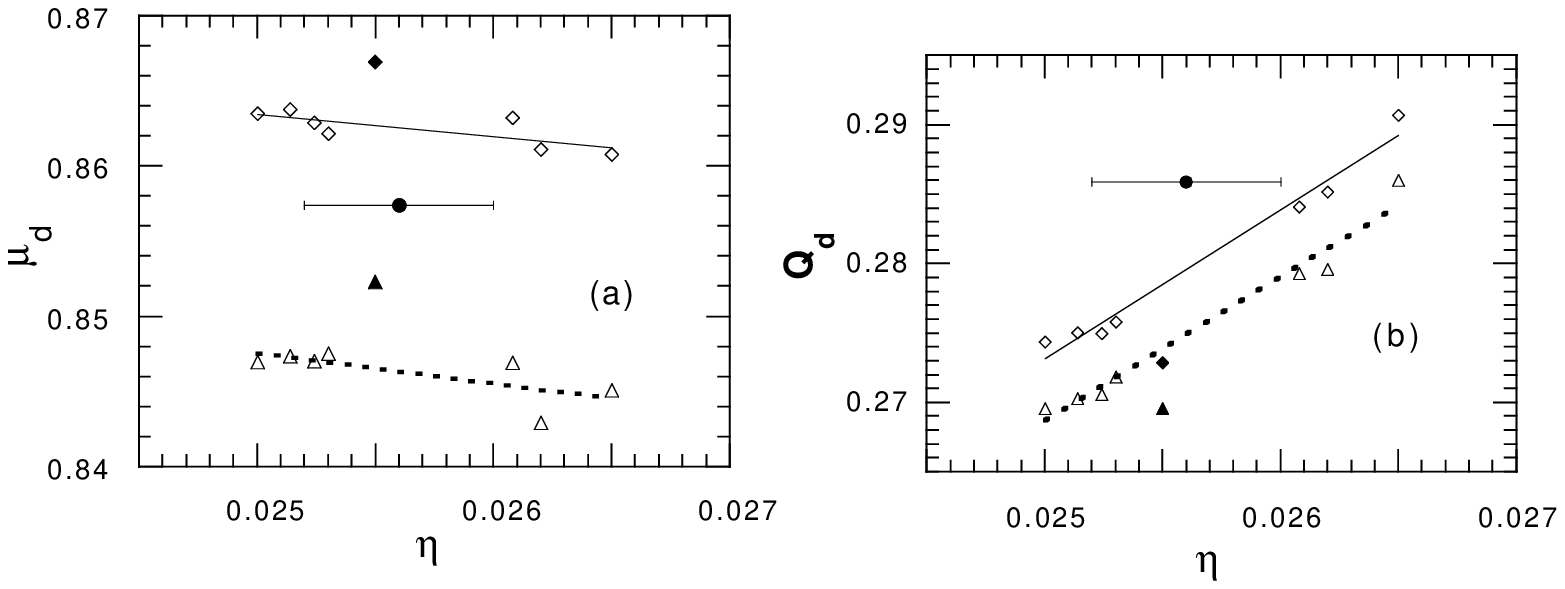,bbllx=10mm,bblly=210mm,bburx=0mm,bbury=282mm}
Figure 1. {(a) Deuteron magnetic moment, $\mu_d$, against the
asymptotic normalization ratio $\eta$, for the Av14
\cite{AV14}, RSC \cite{RSC}, Paris \cite{Paris}, CD-Bonn \cite{CDB}, Nijmegen1,
Nijmegen93, RSC93 \cite{Nij}, and
Av18 \cite{AV18} $N-N$ interactions (mentioned in decreasing order for
$\eta$).  Full dot represents the experimental values for $\mu_d$ and
$\eta$; empty triangles and diamonds correspond to the non-relativistic and
relativistic results, respectively, while dashed and solid lines are linear
fits for these results. Full triangle and diamond are the results of the CD-Bonn
interaction. (b) The same as in (a), but for the deuteron quadrupole moment,
$Q_d$. (After Ref. \cite{LPS}(e))}       
\end{figure} 
\begin{figure}[t]
\psfig{figure=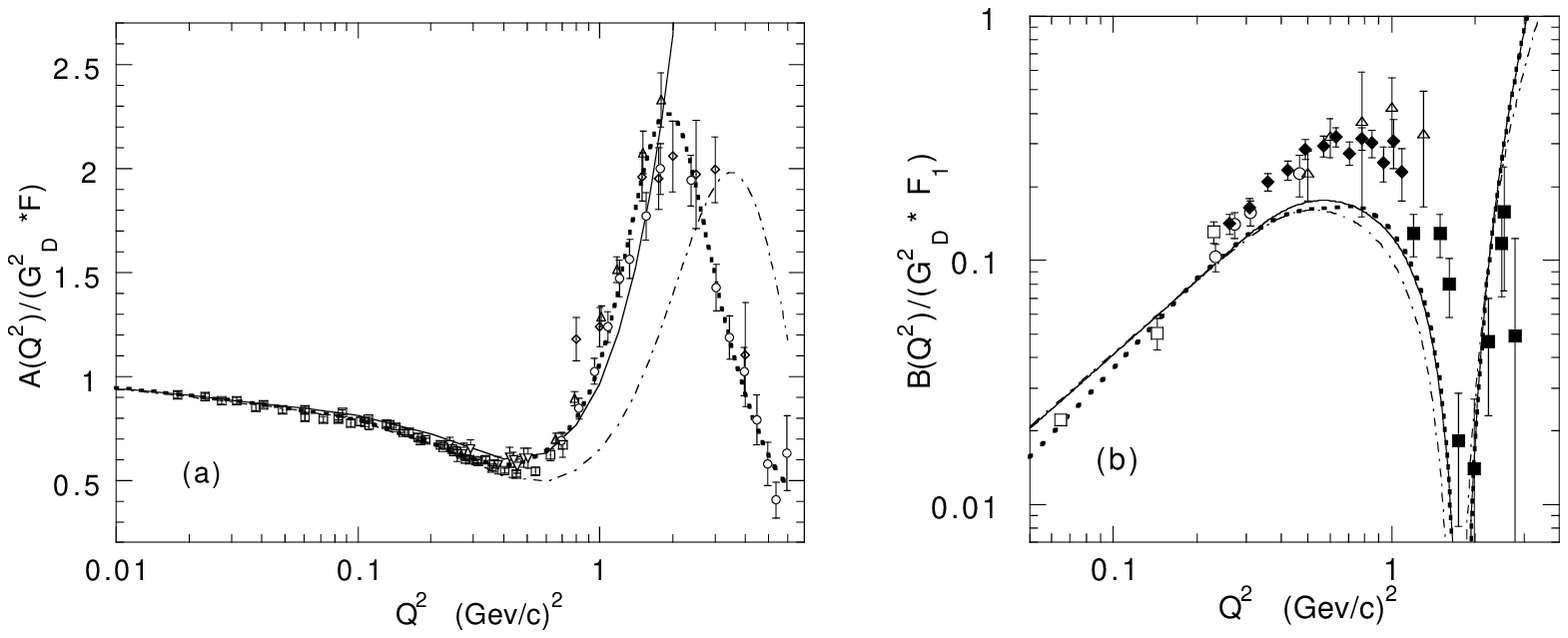,bbllx=10mm,bblly=210mm,bburx=0mm,bbury=282mm}
Figure 2. {(a) The deuteron form factor $A(Q^2)/(G_D^2*F)$ (with $G_D = (1+Q^2/0.71)^{-2}$ and $F = (1+Q^2/0.1)^{-2.5}$)
obtained by the $N-N$ Reid soft core interaction \cite{RSC} and the nucleon
f.f. of Ref. \cite{Gari} (solid line), and Ref. \cite{Hoehler}
(dot-dashed line). The dotted line is obtained using the proton f.f. of
Ref. \cite{Gari} and a fit for the neutron form factors. For the experimental data see
Refs. \cite{LPS}(c,d). (b) The same as
in (a), but for deuteron form factor $B(Q^2)/(G_D^2*F_1)$ (with $F_1 = (1+Q^2/0.1)^{-3}$).}
\end{figure}

In Fig. 2 we report our results for the deuteron f.f. $A(Q^2)$ and $B(Q^2)$. For
$A(Q^2)$ the dependence on the nucleon f.f. is very high,
stronger than the effect of different $N - N$ interactions (see also Refs.
\cite{LPS}(c,d)).  If the poorly known neutron electromagnetic structure is
properly fitted, the overall behaviour of the experimental data can be
reproduced, in particular the recent $A(Q^2)$ data of Refs. \cite{JLABA} and
\cite{JLABC} (open dots and upward triangles, respectively, in Fig. 2(a)) are well described. With respect
to the nucleon f.f. model of Ref. \cite{Gari}, small changes are obtained for
$G_M^n$ and higher values for $G_E^n$ in the range $0.5 - 1 (GeV/c)^2$. The
tensor polarization $T_{20}(Q^2)$ has a considerable dependence on the
interaction \cite{LPS}(c,d) and only a weak dependence on the nucleon f.f., as well known, so
that it cannot be described, together with $A(Q^2)$ and $B(Q^2)$, by a simple fit of the
neutron form factors. In order to obtain a more precise description of the data one
should explicitly introduce two-body currents, which will have to fulfill
separately the constraints of extended Poincar\'e covariance and hermiticity.  A
more stringent comparison with new TJNAF data for $B(Q^2)$ will be possible in
the near future.

\section{CONCLUSION}
\label{S3}
 In the Breit frame where the three-momentum transfer is directed along the spin
quantization axis, the current operator has to be covariant for
rotations around the $z$ axis.
All the necessary
requirements for extended Poincar\'e covariance, hermiticity, current
conservation and charge normalization can be satisfied by a current
obtained by free, one-body terms in that frame \cite{LPS}(a,e). 

   We have applied our results to the calculation of the deuteron elastic f.f.,
without any ambiguity. We were able to obtain, for the first time in a covariant
relativistic approach without {\em{ad hoc}} assumptions on the values of specific matrix elements of the current, unambiguos results for both $\mu_d$ and $Q_d$, close to
the experimental data \cite{LPS}(e). The f.f. $A(Q^2)$ and $B(Q^2)$ show
remarkable effects from different models for the nucleon f.f., while
$T_{20}(Q^2)$ has an higher dependence on different $N-N$ interactions.

 Our approach, based on the reduction of the whole complexity of the Poincar\'e
covariance to the SU(2) symmetry \cite{LPS}(a), can
 represent a simple framework where to investigate  the many-body 
 terms to be added to the free current.

\end{document}